\documentclass[aps,prl,showpacs,twocolumn,groupedaddress]{revtex4}  
\usepackage{graphicx}  

\usepackage{dcolumn}   

\usepackage{bm}        

\usepackage{amssymb}   

\begin{document}
\title{Erratum to  Measurement of $\bm{\sigma (p \bar p \rightarrow Z)
\cdot}$Br$\bm{(Z \rightarrow \tau \tau)}$  at $\bm{\sqrt{s}=}$1.96 TeV,
 published in  Phys. Rev. D {71}, 072004 (2005)}
\date{\today}

\maketitle

 The measurement of ${\sigma (p \bar p \rightarrow Z)
\cdot}$Br${(Z \rightarrow \tau \tau)}$ published in 2005 \cite{D0:Ztt}
  requires a correction for an increase in the reported integrated luminosity.
The instantaneous luminosity at D\O ~is measured by counting the number of
  inelastic collisions that produce charged particles within the acceptance of
  the luminosity monitor \cite{D0:detector}. The determination of the 
luminosity  has recently been improved through studies of the multiplicities 
observed in the luminosity monitor~\cite{D0:lumi_updated}. These studies 
indicated that the fraction of observable inelastic collisions was 
overestimated in our previous analysis~\cite{D0:lumi_old}. For this analysis 
the estimated integrated luminosity increased from 226~pb$^{-1}$
to 257~pb$^{-1}$ and the luminosity uncertainty decreased from $6.5\%$ to $6.1\%$.
The corrected value for ${\sigma (p \bar p \rightarrow Z)
\cdot}$Br${(Z \rightarrow \tau \tau)}$ is 
$209\pm 13 {\rm (stat.)}\pm16 {\rm (syst.)} \pm 13 {\rm (lum.)}$~pb. 
The new value is in 
reasonable agreement with the NNLO standard model predictions of 
$252^{+5}_{-12}$~pb  using the MRST2004 and $242^{+3.6}_{-3.2}$~pb using the 
CTEQ6.1M parametrization of the parton distributions~\cite{TH}.

\begin{thebibliography}{99}
%
\bibitem{D0:Ztt} V.M. Abazov {\sl et al.} (D\O ~Collaboration),
 Phys. Rev. D {\bf 71}, 072004 (2005).
\bibitem{D0:detector}  V.M. Abazov {\sl et al.} (D\O ~Collaboration),
  Nucl. Instrum. Methods Phys. Res. A {\bf 565}, 463 (2006). 
\bibitem{D0:lumi_updated} T. Andeen {\sl et al.}, FERMILAB-TM-2365 (2007).
\bibitem{D0:lumi_old} T. Edwards {\sl et al.}, FERMILAB-TM-2278-E (2004).
\bibitem{TH}
R.~Hamberg, W.L.~van Neerven and T.~Matsuura,
Nucl. Phys. {\bf B359} 343 (1991) [Erratum-ibid. {\bf B644} 403 (2002)].
Additional description of uncertainties can be found in:
A.D.~Martin, R.G. Roberts, W.J.~Stirling and R.S.~Thorne,
Eur. Phys. J. C {\bf 35}, 325 (2004);
A.D.~Martin, R.G. Roberts, W.J.~Stirling and R.S.~Thorne,
Phys. Lett. B {\bf 604}, 61 (2004);
J. Pumplin {\sl et al.}, JHEP {\bf 0207}, 012 (2002);
D. Stump {\sl et al.}, JHEP {\bf 0310}, 046 (2003).

%
\end{thebibliography}
\end{document}